\documentclass[12pt]{revtex4}
\usepackage[latin9]{inputenc}
\usepackage{textcomp}
\usepackage{amsmath}
\usepackage{bm}
\usepackage[toc,page]{appendix}
\usepackage{graphicx}   
\usepackage{verbatim}   
\makeatletter
\@ifundefined{textcolor}{}
{%
 \definecolor{BLACK}{gray}{0}
 \definecolor{WHITE}{gray}{1}
 \definecolor{RED}{rgb}{1,0,0}
 \definecolor{GREEN}{rgb}{0,1,0}
 \definecolor{BLUE}{rgb}{0,0,1}
 \definecolor{CYAN}{cmyk}{1,0,0,0}
 \definecolor{MAGENTA}{cmyk}{0,1,0,0}
 \definecolor{YELLOW}{cmyk}{0,0,1,0}
}

\usepackage{textcomp}
\usepackage{amstext}
\usepackage{amsfonts}
\usepackage{amssymb}
\usepackage{esint}
\def\beq{\begin{equation}}
\def\eeq{\end{equation}}
\def\beqn{\begin{eqnarray}}
\def\eeqn{\end{eqnarray}}
\global\long\def\la{\lambda}

\global\long\def\si{\sigma}

\makeatother

\begin{document}

\title{On instability of ground states in 2D \boldmath{$\mathbb{CP}^{\,N-1}$\! and $\mathbb{O}^{N}$\! models\\ at large $N$}}

\author{A.\ Gorsky,$^{1,2}$ A.\ Pikalov,$^{2,3}$ and A.\ Vainshtein$^{\,4,5}$}

\affiliation{ $^{1}$Institute of Information Transmission Problems of the Russian
Academy of Sciences,\\ Moscow, Russia,\\ 
 $^{2}$\mbox{Moscow Institute of Physics and Technology, Dolgoprudny 141700,
Russia} \\
 $^{3}$Institute for Theoretical and Experimental Physics, Moscow,
Russia\\
 $^{4}$School of Physics and Astronomy and William I.\ Fine Theoretical Physics Institute, University of Minnesota, Minneapolis, USA\\
 $^{5}$\mbox{Kavli Institute for Theoretical Physics, University of California,} Santa-Barbara, USA}

\begin{abstract}
~\\[1.5cm]
We consider properties of the inhomogeneous solution found recently
for \mbox{$\mathbb{CP}^{\,N-1}$} model. The solution was interpreted as a soliton. 
We  reevaluate its energy in three different ways and find
that it is negative contrary to the previous claims. Hence, instead of the solitonic interpretation 
it calls for reconsideration of the issue of the true ground state.
While complete resolution is still absent we show that the energy density
of the periodic elliptic solution is lower than the energy density
of the homogeneous ground state. We also discuss similar solutions for
the ${\mathbb{O}}(N)$ model and for SUSY extensions. 
\end{abstract}
\maketitle

\newpage
\tableofcontents

\section{Introduction}

Two-dimensional $\mathbb{CP}^{\,N-1}$ sigma-model allows exact solution at large
$N$ \cite{dada,witten} and represents such nonperturbative effects
as gap generation, condensates, nontrivial $\theta$-dependence.
It is an asymptotically free theory and in many respects serves as the
laboratory for investigation of  complicated nonperturbative phenomena
in QCD \cite{nsvz}. It was usually assumed that in the infinite volume
the theory is in the confinement phase. However, more recently it was
demonstrated that the phase transition from the confinement phase
to the Higgs phase occurs if the model is perturbed by the twisted
mass term \cite{ferrari,gsy}, considered on $S^{1}$ \cite{unsal} or at
the finite interval \cite{milekhin,Bolognesi}. 

It was known for a while
that in spite of many similar properties of 2D $\mathbb{CP}^{\,N-1}$ and QCD there is one notable
difference -- the signs of the nonperturbative vacuum
energies in 2D $\mathbb{CP}^{\,N-1}$ sigma-model and QCD are opposite \cite{nsvz}.
In QCD the vacuum energy density is proportional to the gluon condensate,
\begin{equation}
\epsilon_{vac}^{\,{\rm QCD}}=\frac 1 4 \langle \theta_{\mu}^{\mu}\rangle =\Big\langle M\,\frac{d{\cal L}^{\,{\rm QCD}}}{dM}\Big\rangle
=\frac{1}{32g^{4}}\,M\,\frac{d g(M)}{dM} \,\big\langle {\rm Tr}\, G^{\mu\nu}G_{\mu\nu}\big\rangle\,,
\end{equation}
while in $\mathbb{CP}^{\,N-1}$ it is the $\langle - D_{\mu}\bar n_{a} D^{\mu} n^{a}\rangle$
condensate instead,
\begin{equation}
\epsilon_{vac}^{{\,\rm CP}}=\frac 1 2 \langle \theta_{\mu}^{\mu}\rangle=
\Big\langle M\,\frac{d{\cal L}^{\,{\rm CP}}}{dM}\Big\rangle
=\frac{1}{4g^{4}}\,M\,\frac{d g(M)}{dM} \,\langle - D_{\mu}\bar n_{a} D^{\mu} n^{a}\rangle\,.
\end{equation}
Both theories are asymptotically free, i.e. have $Mdg/dM < 0$, and  both condensates
$\,\big\langle {\rm Tr}\, G^{\mu\nu}G_{\mu\nu}\big\rangle$ and $\langle - D_{\mu}\bar n_{a} D^{\mu} n^{a}\rangle$
are positively definite in the Eucledean signature. However, the gluon condensate is positive in its both perturbative 
and nonperturbative pieces while positivity of $\langle - D_{\mu}\bar n_{a} D^{\mu} n^{a}\rangle$ is due to 
perturbative part only -- nonperturbative part is negative, see \cite{nsvz} for details.

The model can be also considered in the SUSY setting and it turns
out that the observed similarity between $\mathbb{CP}^{\,N-1}$ model and QCD has
very attractive explanation in the SUSY context. The SQCD allows the
non-abelian strings \cite{tong,konishi,sy} and the SUSY--$\mathbb{CP}^{\,N-1}$
is just the world-sheet theory on the non-abelian string (see \cite{rev1,rev2}
for the review). The degrees of freedom in  $\mathbb{CP}^{\,N-1}$ model are identified
with the orientational modes on the non-abelian string. A similar
non-abelian string solution occurs also in the non-SUSY 4D gauge model 
which is essentially the bosonic part of the SQCD Lagrangian \cite{gsy2}. In this case
the worldsheet theory on the string is non-SUSY   $\mathbb{CP}^{\,N-1}$ model.

There is 2D-4D correspondence \cite{dorey} between SQCD
and the world-sheet theory on the defect. It claims that running
of the coupling constant, spectrum of the stable particles, twisted
superpotentials in 4D and 2D theories fit each other.
The very 2D-4D correspondence reflects the property
that the non-abelian string can exist on the top of the SQCD vacuum not
destroying it as the electron can propagate at the top of the
Cooper condensate. It just makes quantitative that properties of any object 
considered from the viewpoints of 2D and 4D observers should be the same.

Recently the new inhomogeneous solution to  $\mathbb{CP}^{\,N-1}$ model has been
found in Ref.\,\cite{Nitta}. The key tool for the derivation of the solution
was the particular mapping of the $\mathbb{CP}^{\,N-1}$ model to the Gross-Neveu (GN)
model. The new solution of $\mathbb{CP}^{\,N-1}$ model was obtained from the kink solution
of GN model interpolating between two vacua with the different
values of the fermion condensate. More general kink lattice
configuration has been found as well using the elliptic solution to
the GN model. This inhomogeneous solution and especially the lattice
solution has some common properties with the inhomogeneous condensates
in the GN and the chiral GN models \cite{dunne1,dunne2}.  Note
that there is also some analogy with the Peierls model of 1+1 superconductivity.
It that case the electron propagates along some nontrivial profile
of the lattice state and the integrability of the model allows to
get its exact solution in some continuum \cite{peierls} and discrete 
cases \cite{discrete}. The fermions
play the role of the eigenfunctions for the Lax operator for some
integrable model  and the spectral curve describing
the finite-gap solution simultaneously plays the role of the dispersion
law for the fermions. The ground state of the system strongly depends
on the fermionic density and the temperature.

In this study we focus at some aspects of this new solution. 
We reevaluate accurately its energy and find that it is negative contrary
to the statement made in \cite{Nitta}. Three different approaches
of derivation of the ground state energy yield the same result.
This raises the question concerning the true ground state of the model.
We shall argue that the inhomogeneous solution and in particular 
the elliptic soliton lattice are the candidate  ground state 
of the model. However, there are some reservations due to the 
IR properties of the solution.

Let us recall that  the conventional viewpoint implies an existence of
single homogeneous ground state separated by the small gaps of order
${1}/{N}$ from the set of the metastable vacua. The ground state of $\mathbb{CP}^{\,1}$ model
becomes degenerate only at $\theta=\pi$ when kinks are allowed, and in SUSY case for  $\mathbb{CP}^{N\!-\!1}$  when $N$ degenerate
vacua exist. At one loop-level the kinetic
term for the photon is generated which yields the
linear potential between  charges. It was argued in \cite{witten} 
that the excitations of the 
model are identified as  the singlet $n^{*}n$ states. It was also noted in \cite{witten}
that the n-particle corresponds to the kink between two 
vacua if the fermions are added to the Lagrangian.  To some extent $n^{*}n$ pair
corresponds to the interpolation between the excited metastable
vacuum and the true one. In this paper we question this standard picture. 

The soliton solution in the $\mathbb{CP}^{\,N-1}$ model  obtained in \cite{Nitta}
is the counterpart of the elementary kink solution in the GN model or the composite 
kink solution in the chiral GN model.
In the GN model there are two vacua therefore the interpolating kink
with the well-defined topological charge does exist. The topology guarantees
its stability. Since it is this solution which gets mapped into $\mathbb{CP}^{\,N-1}$
solution we could wonder if there is some topological reason which
yields the stability of new solution in $\mathbb{CP}^{\,N-1}$ case. 

We also discuss the similar solution in ${\mathbb O}^{N}$ model and in the $\cal{N}$=1 SUSY
extensions. Although the kinks in SUSY case are well defined
BPS particles saturating the corresponding central charges the evaluation
of their masses was the controversial issue for a while with several
different answers. This puzzle has been resolved in \cite{vai1,vai2}
where the effect of anomalies has been taken into account carefully.
The finite effects of the anomalies in the mode counting has been
also found in the non-SUSY $\mathbb{CP}^{\,N-1}$ model in \cite{bolognesi2}.

The paper is organized as follows. In Section II we recall the main features
of the nonperturbative solution to the $\mathbb{CP}^{\,N-1}$  model and  the inhomogeneous solution
is derived via the method of resolvent. Its energy is evaluated
by three different approaches in Section III. Some remarks concerning the SUSY generalization
of the solution are presented in Section IV while the elliptic kink crystal
solution is considered in Section V. The results and open questions are summarized
in the Discussion, Secton VI, while some technical details are collected in the Appendices.

\section{\boldmath{$\mathbb{CP}^{\,N-1}$} model}

\subsection{Saddle point equations}
Let us remind the standard derivation of the saddle point approximation to the solution.
Lagrangian of $\mathbb{CP}^{\,N-1}$ model in Minkowski space is 
\begin{equation}
\mathcal{L}=D^{\mu}\bar n_{a}\,D_{\mu}n^{a}-\lambda\, (\bar n_{a} n^{a}-r)\label{eq:cp}
\end{equation}
where $n^{a},\;a=1,\dots,\,N$ are complex fields in the fundamental representation of SU($N$),
\mbox{$r=1/g^{2}$} defines the coupling constant,
$\bar n_{a}=(n^{a})^{*}$ and $\lambda$ is the Lagrange
multiplier. Moreover, $D_{\mu}n^{a}=(\partial_{\mu}+iA_{\mu})\,n^{a}$ where $A_{\mu}$ is a dummy
field. 

Let us go to Euclidian signature and integrate over $N-1$ fields $n^{a},\;a=1,\dots,\,N-1$, but not over $n^{N}=n$.
Due to gauge invariance the $n^{N}$ field can be chosen to be real. Besides the field $n$ the arising effective action depends on two more real fields:  $\lambda$ and $A_{\mu}$. For $A_{\mu}=0$ the Euclidian effective action takes the form
\begin{equation}
S=(N-1){\rm Tr}\log\left(-\partial^{2}+\lambda\right)+\int d^{2}x\left((\partial n)^{2}+\lambda\left(n^{2}-r\right)\right)
\label{eq:cp action}
\end{equation}

Let us write now the saddle point equation implying that the fields $\lambda$ and $n$ are static, i.e., do not depend on time,
but could depend on space coordinate $x$. Variation over $n(x)$ leads to
\begin{equation}
\left(\partial_{x}^{2}-\lambda\left(x\right)\right)n\left(x\right)=0 \,,
\label{gap2}
\end{equation}
what allows to express $\lambda$ in terms of $n$,
\begin{equation}
\lambda=\frac{\partial_{x}^{2}n}{n}\,.
\label{gap3}
\end{equation}
From variation over $\lambda(x)$ we get (neglecting difference between $N-1$ and $N$),
\begin{equation}
\int dt\left[N\big\langle x,t\big|\frac{1}{-\partial_{t}^{2}-\partial_{x}^{2}+\lambda}\big|x,t\big\rangle+n^{2}-r\right]=0\,,
\end{equation}
what is equivalent to
\begin{equation}
\frac{N}{2\pi}\int d\omega\,\big\langle x\big|\frac{1}{-\partial_{x}^{2}+\omega^{2}+\lambda}\big|x\big\rangle+n^{2}\left(x\right)-r=0\label{gap}.
\end{equation}
For the homogenous solution with $\lambda=m^{2}$ the field $n=0$
and 
\begin{equation}
r=\frac{N}{\left(2\pi\right)^{2}}\int d\omega dk\,\frac{1}{k^{2}+\omega^{2}+\lambda}=\frac{N}{4\pi}\int d\omega\,\frac{1}{\sqrt{\omega^{2}+m^{2}}}=\frac{N}{2\pi}\log\frac{M}{m}\,,
\end{equation}
where $M$ denotes the UV cut-off introduced via Pauli-Villars regularization (see part B in Sec. III for details).

For inhomogeneous solution we can then rewrite Eq.\,(\ref{gap}) as
\begin{equation}
n^{2}\left(x\right)=\frac{N}{2\pi}\int_{-\infty}^{\infty}d\omega\left[\frac{1}{2\sqrt{\omega^{2}+m^{2}}}-R_{\omega}\left(x\right)\right]
\,,
\label{gap1}
\end{equation}
where  $R_{\omega}$ denotes the resolvent,
\begin{equation}
R_{\omega}=\Big\langle x\Big|\frac{1}{-\partial_{x}^{2}+\omega^{2}+\lambda}\Big|x\Big\rangle.
\end{equation}
The equation (\ref{gap1}) can be also written as a sum over eigenfunctions of
the operator $-\partial_{x}^{2}+\lambda$,
\begin{equation}
n^{2}=r-N\sum\frac{\left|f_{k}\left(x\right)\right|^{2}}{2\omega_{k}},\qquad
(-\partial_{x}^{2}+\lambda(x))f_{k}(x)=\omega_{k}^{2}\,f_{k}(x)\,.
\label{eq:gap1 eigenfunctions}
\end{equation}

In finding a inhomogeneous solution the main
idea is to use well-known fact that resolvent $R_{\omega}$ satisfies
the Gelfand-Dikii equation 
\begin{equation}
-2R_{\omega}\partial_{x}^{2}R_{\omega}+\left(\partial_{x}R_{\omega}\right)^{2}+4\left(\omega^{2}+\lambda(x)\right)R_{\omega}^{2}=1\label{eq:gd}
\end{equation}
If we use the relation (\ref{gap3}) to substitute $\lambda$ 
and propose some ansatz for $R_{\omega}$ we obtain a differential
equation for $n$ with parameter $\omega$. This equation must
hold for all values of $\omega$ which is possible only for special
choice of coefficients.

Assume that the spectrum of
Schrodinger operator  consists
of one translational zero mode and continuum starting at eigenvalue
$\omega^{2}=m^{2}$. 
Hence we suppose that 
\begin{equation}
R_{\omega}=a\left(\omega\right)+b\left(\omega\right)n^{2}\left(x\right)\label{ansatz}
\end{equation}
This is the simplest choice which is consistent with (\ref{gap1}).
It is also reasonable to assume that 
\begin{equation}
a\left(\omega\right)=\frac{1}{2\sqrt{\omega^{2}+m^{2}}}\label{assume a}
\end{equation}
but for a moment we will not use this assumption. After substitution
of (\ref{ansatz}) and (\ref{gap2}) in (\ref{eq:gd}) we obtain the
equation 
\begin{equation}
4a\left(a+bn^{2}\right)\partial_{x}^{2}n+4\omega^{2}n\left(a+bn^{2}\right)^{2}-4abn\left(\partial_{x}n\right)^{2}=n\label{main eq}
\end{equation}

If we use (\ref{assume a}) and assume $b=Ca/\omega^{2}$ where $C$
is some constant we obtain that (\ref{main eq}) is equivalent to
two equations 
\begin{equation}
n\partial_{x}^{2}n+Cn^{4}-\left(\partial_{x}n\right)^{2}=0\label{eq1}
\end{equation}
\begin{equation}
\partial_{x}^{2}n+2Cn^{3}=m^{2}n\label{eq2}
\end{equation}
From these equations we easily obtain that
\begin{equation}
\left(\partial_{x}n\right)^{2}=n^{2}\left(m^{2}-Cn^{2}\right)
\end{equation}
For $C>0$ the solution is 
\begin{equation}
n\left(x\right)=\frac{m}{\sqrt{C}}\,\frac{1}{\cosh\left(m\left(x-x_{0}\right)\right)}
\end{equation}
where $x_{0}$ is the center of the soliton. Thus, the condensate $\lambda$ is
\begin{equation}
\lambda\left(x\right)=\frac{\partial_{x}^{2}n}{n}=m^{2}\left[1-\frac{2}{\cosh^{2}\left(m\left(x-x_{0}\right)\right)}\right]\label{eq:line cond}
\end{equation}
This is the solution found in \cite{Nitta}. Eigenfunctions with given momentum at infinity
may be found via supersymmetric quantum mechanics,
\begin{equation}
\begin{split}
&\left(-\partial_{x}^{2}+\lambda\left(x\right)\right)f_{k}\left(x\right)=\omega_{k}^{2}f_{k}\left(x\right),\\
&\omega_{k}^{2}=m^{2}+k^{2}\,,\qquad
f_{k}\left(x\right)=\frac{-ik+m\tanh mx}{\sqrt{m^{2}+k^{2}}}\,\exp\left(ikx\right)\,.
\label{eq:soliton eigenfunctions}
\end{split}
\end{equation}
We put $x_{0}=0$ above.
These functions are normalized as 
\[
\int_{-\infty}^{+\infty}dxf_{k}\left(x\right)f_{k^{\prime}}^{*}\left(x\right)=2\pi\delta\left(k-k^{\prime}\right)
\]
Thus,  from Eq.\,(\ref{eq:gap1 eigenfunctions}) we get the same solution,
\begin{equation}
n^{2}(x)\!=\!N\!\int \!\frac{dk}{2\pi}\left[\frac{1}{2\sqrt{k^{2}+m^{2}}}-\frac{\left|f_{k}\left(x\right)\right|^{2}}{2\sqrt{k^{2}+m^{2}}}\right]=\frac{N}{4\pi}\!\int\! dk\frac{m^{2}\left(1\!-\!\tanh^{2}mx\right)}{\left(k^{2}+m^{2}\right)^{3/2}}=\frac{N}{2\pi}\frac{1}{\cosh^{2}mx}\,.
\end{equation}

Let us comment on the topological aspect of the solution. In the GN model
the kink interpolates between two vacuum states and has the  standard
topological charge which is due to the difference of the field at two spatial
infinities. Our soliton has no  naive local topological charge since  
values of the fields at two space asymptotics are the same. The solution
looks like the soliton solution in the KdV equation and in the 
integrability context one could say that selecting the soliton 
solution which has positive energy we select the topological sector of the theory and the
topology can be read off only from the geometry of the spectral curve.

In our case if our solution would have the conserved topological charge and have 
the positive energy one could claim that it is just particular sector
of excitations above the ground state. However there is no local conserved
charge and its energy is negative hence we interpret it as the instability
mode for the homogeneous ground state.

\section{Energy of the soliton}

In this Section we will provide three different ways of evaluation of energy
for the solution obtained in the previous section. Firstly we will
use simple regularization by introducing ultraviolet cut-off and taking
into account the anomaly found in \cite{bolognesi2}. Then we obtain
the same result using Pauli-Villars regularization. Finally, we
calculate the average of energy-momentum tensor. A bit surprisingly
in all calculations we obtain a negative value for the soliton energy 
\begin{equation}
E=-\frac{2Nm}{\pi}\label{eq:energy}
\end{equation}

\subsection{Regularized sum over the modes}

We first use the expression from \cite{bolognesi2} that energy density
for a static configuration of $\lambda$ which satisfies the gap equation
is 
\begin{equation}
\varepsilon\left(x\right)=\varepsilon_{0}+\frac{N}{2\pi}\,\lambda\left(x\right),\qquad \partial_{x}\varepsilon_{0}=0\,.
\end{equation}
Let us emphasize that this expression
takes into account the anomalous contribution emerging from the regularization
of the sum over the modes.

If we subtract the vacuum energy density $\varepsilon_{vac}$ given the same expression with $\lambda=m^{2}$
we obtain 
\[
\varepsilon\left(x\right)-\varepsilon_{vac}=const+\frac{N}{2\pi}\left(\lambda\left(x\right)-m^{2}\right).
\]
It is reasonable to assume that at spacial infinity energy density
is the same as in vacuum so $const=0$. After substitution of solution
(\ref{eq:line cond}) into the energy density and integration we find
\begin{equation}
E=\int_{-\infty}^{+\infty}dx\left(\varepsilon\left(x\right)-\varepsilon_{vac}\right)=
-\frac{Nm^{2}}{\pi}\int_{-\infty}^{+\infty}\!\!dx\,\frac{1}{\cosh^{2}mx}=-\frac{2Nm}{\pi}\,.
\end{equation}

Since the energy of the soliton derived in \cite{Nitta} is different
and positive one could wonder what is the reason for the discrepancy.
In \cite{Nitta} the following expression for the energy was used
$E=N\sum\omega_{n}-r\int dx\lambda+b.t.$ and the derived energy of
soliton is positive and reads as $E_{sol}-E_{0}=r\int(\lambda_{o}-\lambda_{sol})=4rm$
where the complete cancellation of the sum over the modes around the
vacuum and soliton was assumed. The first point of concern is the
presence of the bare coupling constant $r$ in the expression for
the quantum energy. The second point which is not correct is the complete
cancellation of the modes at the top of the solution which was shown to be incomplete \cite{vai1,vai2}.
Finally the anomaly for the energy due to the proper regularization
procedure \cite{bolognesi2} has not been taken into account.

\subsection{Pauli-Villars regularization}

We calculate energy of the soliton by regularizing its effective action
by Pauli-Villars method. In this calculation we follow ideas from
\cite{nsvz}. The regularized action is 
\[
S=N\sum_{i=0}^{I}C_{i}{\rm Tr}\log\left(-\partial^{2}+m_{i}^{2}+\lambda\right)+
\!\int\! d^{2}x\left[(\partial n)^{2}+\lambda\left(n^{2}-r\right)\right]
\]
Following the Pauli - Villars procedure, we introduce in addition to each original field with $m_{0}=0$ a number $I$ 
of regulator fields with masses $m_{i},~i=1,\dots,\,I$, and constants $C_{i},~i=0,1,\dots,\,I$,
satisfying 
\[
\sum_{i=0}^{I}C_{i}=0\,, \quad \sum_{i=0}^{I}C_{i}m_{i}^{2}=0\,,\quad C_{0}=1\,, \quad m_{0}=0\,.
\]
For our purposes it is sufficient to take $I=2$. Then 
the constants $C_{i}$ are 
\[
C_{1}=\frac{m_{2}^{2}}{m_{1}^{2}-m_{2}^{2}}\,,\quad C_{2}=-\frac{m_{1}^{2}}{m_{1}^{2}-m_{2}^{2}}\,.
\]
At the end of calculation we will take a limit when all regulator masses $m_{i}\,(i=1,\dots,\,I)$ go
to the UV cut-off $M$.
The connection between effective action and energy is $S=E\cdot T$,
where $T$ is a large time cut-off.

The general scheme of calculations is as follows. First, we find coupling
constant $r$ in terms of regulator fields masses and mass scale of
the theory from the gap equation for homogeneous solution $\lambda=m^{2}$.
Next, we can show that terms with the $n$ field do not contribute
to the energy because $n$ is proportional to zero mode:
\[
\int_{-\infty}^{+\infty}dx\left[\left(\partial_{x}n\right)^{2}+\lambda\, n\right]=\int_{-\infty}^{+\infty}\!\!dx\,n\left(-\partial_{x}^{2}n+
\lambda\, n\right)=0\,.
\]
After that we express the trace term of as a sum over eigenvalues
and take into account the change in the density of states for inhomogeneous
solution. Finally, we perform integration over eigenvalues and confirm
the result \eqref{eq:energy}. Details of the computation are presented
in Appendix \ref{sec:Effective-action-calculation}.

\subsection{Energy of soliton, explicit evaluation}

In this section we are going to calculate the average of energy-momentum
tensor for a soliton solution. We quantize the $n$ fields canonically
and introduce Pauli-Villars regulator fields to deal with divergences
and take into account the conformal anomaly. The energy momentum tensor
in Minkowski space is 
\[
\theta_{\mu\nu}=\sum C_{i}\theta_{\mu\nu}^{i}\,,\quad \theta_{\mu\nu}^{i}=\partial_{\mu}n_{i}\partial_{\nu}n_{i}^{*}+\partial_{\mu}n_{i}^{*}\partial_{\nu}n_{i}-g_{\mu\nu}\left(\left|\partial n_{i}\right|^{2}-\lambda\left(\left|n_{i}\right|^{2}-r\right)-m_{i}^{2}\left|n_{i}\right|^{2}\right).
\]
The components $\theta_{00}\,,\theta_{11}$ are 
\begin{equation}
\begin{split}
&\theta_{00}=\sum C_{i}\left(\left|\partial_{t}n_{i}\right|^{2}+\left|\partial_{x}n_{i}\right|^{2}+\lambda\left|n_{i}\right|^{2}+m_{i}^{2}\left|n_{i}\right|^{2}\right)-\lambda r\,,\\
&\theta_{11}=\sum C_{i}\left(\left|\partial_{t}n_{i}\right|^{2}+\left|\partial_{x}n_{i}\right|^{2}-\lambda\left|n_{i}\right|^{2}-m_{i}^{2}\left|n_{i}\right|^{2}\right)+\lambda r\,,\\
&\theta_{01}=\sum C_{i}\left(\partial_{t}n_{i}\,\partial_{x}n^{*}_{i}+\partial_{t}n_{i}^{*}\,\partial_{x}n_{i}\right).
\end{split}
\label{eq:time energy-momentum}
\end{equation}

We consider field $\lambda$ as classical and suppose that the $n$
field has a classical component:
\[
\lambda=m^{2}\left(1-\frac{2}{\cosh^{2}mx}\right),
\qquad
n_{cl}=\sqrt{\frac{N}{2\pi}}\,\frac{1}{\cosh mx}\,.
\]
The modes on the $n$ field in continuum spectrum are given by Eq.\,\eqref{eq:soliton eigenfunctions}.
Also there is a zero mode
\[
\psi_{0}=\sqrt{\frac{m}{2}}\,\frac{1}{\cosh mx}.
\]
Quantization of field $n=n^{N}$ and regulator fields $n_{i}\,, (i=1,2)$, are slightly
different. The $n$ field has classical component, proportional to zero
mode, while the regulator field have additional component with frequency
$m_{i}$. 
The masses of auxiliary fields and coefficients $C_{i}$ are the same
as in the calculation of the determinant via Pauli-Villars regularization.
The frequencies for regulator fields are $\omega_{k,\,i}=\sqrt{\omega_{k}^{2}+m_{i}^{2}}$.
In terms of creation and annihilation operators we have 
\begin{equation}
n^{a}\left(x,\,t\right)=\delta^{a}_{N}n_{cl}\left(x\right)+\int\frac{dk}{2\pi}\frac{1}{\sqrt{2\omega_{k}}}\left(a_{k}^{a}f_{k}\left(x\right)e^{-i\omega_{k}t}\!+b_{k}^{a\dagger}f_{k}^{*}\left(x\right)e^{+i\omega_{k}t}\right)
\end{equation}
for $n^{a}$ field. For the the regulator fields $n^{a}_{i}\,,\, i=1,...,I,$ we have
\begin{equation}
n_{i}^{a}\! =\! \frac{1}{\sqrt{2m_{i}}}\left(A_{i}^{a}e^{-im_{i}t}\!\!  +\! B_{i}^{a\dagger}e^{+im_{i}t}\right)\psi_{0}\left(x\right)\! 
+ \! \!  \int\! \frac{dk}{2\pi}\frac{1}{\sqrt{2\omega_{k,\,i}}}\!\left(a_{k,\,i}^{a}f_{k}\left(x\right)e^{-i\omega_{k,i}t}\!\!  +\! b_{k,\,i}^{a\dagger}f_{k}^{*}\left(x\right)e^{+i\omega_{k,i}t}\right),
\end{equation}
The canonical commutation relations for $n$ field is modified by the presence of
zero mode:
\[
\left[n^{a}\left(x,\,t\right),\partial_{t}\bar n_{b}\left(y,\,t\right)\right]=i\delta^{a}_{b}\left(\delta\left(x-y\right)-i\delta^{a}_{N}\delta^{N}_{b}\psi_{0}\left(x\right)\psi_{0}\left(y\right)\right).
\]
However, for regulator fields commutation relation is unchanged,
\[
\left[n_{i}^{a}\left(x,\,t\right), \partial_{t}\bar n_{kb}\left(y,\,t\right)\right]=i\delta_{ik}\delta^{a}_{b}\delta\left(x-y\right).
\]

We take average over the state, which is annihilated by all operators
$a_{k}$, $a_{k,i}$, $A_{i}$ and $b_{k}$, $b_{k,i}$, $B_{i}$. For the product of two $n=n^{N}$ fields we get
\[
\langle n\left(x_{1,}t_{1}\right)n^{\dagger}\left(x_{2,}t_{2}\right)\rangle=n_{cl}\left(x_{1}\right)n_{cl}\left(x_{2}\right)+N\int\frac{dk}{2\pi}\frac{1}{2\sqrt{k^{2}+m^{2}}}\,e^{i\omega_{k}(t_{1}-t_{2})}f_{k}^{*}\left(x_{1}\right)f_{k}\left(x_{2}\right).
\]
For corresponding regulators it gives
\[
\langle n_{i}\left(x_{1,}t_{1}\right)n_{i}^{\dagger}\left(x_{2,}t_{2}\right)\rangle\!=\!N\,\frac{\psi_{0}\left(x\right)\psi_{0}\left(y\right)}{2m_{i}}e^{im_{i}(t_{1}-t_{2})}+N\!\!\int\!\frac{dk}{2\pi}\frac{e^{i\omega_{k,i}(t_{1}-t_{2})}}{2\sqrt{k^{2}+m^{2}+m_{i}^{2}}}\,f_{k}^{*}\left(x_{1}\right)f_{k}\left(x_{2}\right).
\]
The expression for the regularized square of the field is then,
\[
\sum_{i=0}^{i=2}C_{i}\langle\left|n_{i}\left(x\right)\right|^{2}\rangle=n_{cl}^{2}\left(x\right)+N\int\frac{dk}{2\pi}\sum_{i}\frac{C_{i}\left|f_{k}\left(x\right)\right|^{2}}{2\sqrt{k^{2}+m^{2}+m_{i}^{2}}}+N\psi_{0}\left(x\right)^{2}\sum_{i}\frac{C_{i}}{2m_{i}}=r\,.
\]
This equality is equivalent to the gap equation, therefore the $r$
term in energy momentum tensor cancels by the $n^{2}$ term.

The calculation of other contributions to energy-momentum tensor is
straightforward. Details are provided in Appendix \ref{sec:Energy-momentum-tensor-of}.
The final answer is consistent with other methods:

\begin{equation}
\big\langle\theta_{00}\big\rangle=\frac{Nm^{2}}{4\pi}-\frac{N}{\pi}\frac{m^{2}}{\cosh^{2}mx}=\frac{Nm^{2}}{4\pi}+\frac{N}{2\pi}\left(\lambda-m^{2}\right).\label{eq: average time}
\end{equation}
The other components of energy-momentum tensor are the same as ones
of the homogeneous phase 
\begin{equation}
\big\langle\theta_{11}\big\rangle=-\frac{Nm^{2}}{4\pi}\,,\qquad \big\langle\theta_{01}\big\rangle=0\,.\label{eq: average space}
\end{equation}

This can be compared with evaluation of the energy density of the
homogeneous ground state via the conformal anomaly \cite{nsvz}. Since
there is no scale at the classical level the trace of the energy stress
tensor gets contribution from the running of the coupling constant
only and therefore is proportional to the $\beta$-function, $\theta_{\mu}^{\,\mu}=N\lambda/2\pi$.
Hence the vacuum energy density  $\epsilon_{vac}=(1/2)\langle vac\,|\,\theta^{\,\mu}_{\mu}\,|vac\rangle=Nm^{2}/4\pi$.
Similarly the mass of the particle can be evaluated from the matrix
element of the $\theta_{\mu}^{\,\mu}$ over the corresponding state \cite{nsvz}.
For instance we can use the relation for the $\sigma$-particle mass,
$2m^{2}=\langle \sigma|\,\theta^{\,\mu}_{\mu}\,|\sigma\rangle$ and express it via the propagator
of the $\lambda$-field $D(p^{2})$ at zero momentum $D(0)$ and simple
$\sigma\sigma\lambda$ vertex proportional to $2m^{2}/N$.

To complete this Section let us make a comment concerning the spectrum
of excitations. First note that the photon acquires finite inhomogeneous
mass in the non-homogeneous vacuum. This implies that there is no 
linear confinement  of  charged degrees of freedom. According to the
emerging picture the homogeneous state is metastable and the kink-antikink
pair in the homogeneous state now yield the bounce configuration in the
Euclidean space. We shall discuss the spectrum 
and the $\theta$-dependence in the inhomogeneous ground state in more details elsewhere.  

\section{\boldmath{${\cal {N}}$=1} Supersymmetric models}

\subsection{SUSY \boldmath{$\mathbb{O}^N$} sigma model}

First let us argue that $\mathbb{O}(N)$ model admits the similar inhomogeneous
solution and then consider its minimal SUSY extension. The Lagrangian
of the model reads as 
\begin{equation}
\mathcal{L}=\frac{1}{2}\left(\partial_{\mu}n_{a}\right)^{2}-\frac{\lambda}{2}\left(\left(n_{a}\right)^{2}-r\right)\label{eq:on}
\end{equation}
There are $N$ real fields $n_{a}$ and Lagrange multiplier $\lambda$
leads to constraint $n_{a}n_{a}=r=1/g^{2}$. Similar to the case of $\mathbb{CP}^{\,N\!-\!1}$
model, this model demonstrates dynamical mass generation, so in vacuum
$\lambda=m^{2}$. It is simple issue to show that in the large $N$ limit
model (\ref{eq:on}) possess a soliton solution similar to the one
being discussed in case of $\mathbb{CP}^{\,N\!-\!1}$ model. The difference
is only in number of degrees of freedom.

Large $N$ effective action is obtained similarly to the case of $\mathbb{CP}^{\,N\!-\!1}$
model by integration over fields $n_{a},\;a=1,2,\dots,\,N-1$, but not over $n_{N}=n$. In the Euclidean signature the effective action is
\begin{equation}
S_{eff}=\frac{N-1}{2}\,{\rm Tr}\log\left(-\partial^{2}+\lambda\right)+\frac{1}{2}\int d^{2}x\left(\left(\partial n\right)^{2}
-\lambda (n^{2}-r)\right).\label{eq:on action}
\end{equation}
The actions (\ref{eq:cp action}) and (\ref{eq:on action}) differ
only by numerical factor of 1/2. Thus, their stationary points are the
same and (\ref{eq:line cond}) is solution in $\mathbb{O}^{N}$ model
with energy 
\[
E=-\frac{Nm}{\pi}\,.
\]

Let us turn now to the case of ${\cal N}=1$ supersymmetric $\mathbb{O}^{N}$
model. The Lagrangian is 
\[
\mathcal{L}=\frac{1}{2}\left[\left(\partial_{\mu}n_{a}\right)^{2}+\bar{\psi}_{a}i{\not\!\partial}\,\psi_{a}+\frac{1}{4\,r}\left(\bar{\psi_{a}}\psi_{a}\right)^{2}\right].
\]
Here $\psi_{a}$ are Majorana fermions, ${\not\!\partial}=\gamma^{\mu}\partial_{\mu}$,
$\gamma^{0}=\sigma_{2},\;\gamma^{1}=i\sigma_{3},\;\gamma^{5}=-\gamma^{0}\gamma^{1}=\sigma_{1}$.
The constraints $n_{a}n_{a}=r$ and $n_{a}\psi_{a}=0$ are taken into
account by Lagrange multipliers $\la$ and $\chi$. Also we introduce
auxiliary field $\sigma\sim\bar{\psi}\psi$,
\[
\mathcal{L}=\frac{1}{2}\left[\left(\partial_{\mu}n_{a}\right)^{2}+\bar{\psi}_{a}\left(i{\not\!\partial}-\sigma\right)\psi_{a}-r\sigma^{2}-\lambda\left(\left(n_{a}\right)^{2}-r\right)-\bar{\chi}\psi_{a}n_{a}-\bar{\psi}_{a}\chi n_{a}\right].
\]
In order to obtain effective action, we have to integrate over all
fermionic fields and all fields $n_{a}$ but $n_{N}=n$. To integrate
over $\psi_{a}$ we make shift of variables 
\[
\psi_{a}\to\psi_{a}+\phi_{a},\quad \phi_{a}=\left(i{\not\!\partial}-\sigma\right)^{-1}\chi n_{a}\,.
\]
Then terms in action linear in $\psi_{a}$ are canceled, but we have
additional term\\ $n_{a}n_{a}\,\bar{\chi}\,(i{\not\!\partial}\!-\!\sigma)^{-1}\chi\!=r\,\bar{\chi}(i{\not\!\partial}\!-\!\sigma)^{-1}\chi$.
Then integration over $\chi$ can also be performed. Integration over
$\psi_{a}$ and $\chi$ yields determinant contributions to effective
action,
\[
-\frac{iN}{2}\,{\rm Tr}\log\left(i{\not\!\partial}-\sigma\right)+\frac{i}{2}\,{\rm Tr}\log\left(i{\not\!\partial}-\sigma\right),
\]
hence, the field $\chi$ inttegration reduces the number of degrees of freedom by 1.
Effective action is 
\begin{equation}
S_{eff}=\frac{i}{2}\,(N\!-\!1)\left[{\rm Tr}\log\left(-\partial^{2}\!-\!\lambda\right)\!-\!{\rm Tr}\log\left(i{\not\!\partial}-\sigma\right)\right]
\!+\!\frac{1}{2}\int\! d^{2}x \left(\left[\partial n\right)^{2}\!-\!\lambda (n^{2}\!-\!r)\!-\!\sigma^{2}r\right]
\label{eq:susy1}
\end{equation}
Note that this action can be rewritten in slightly different way,
making the situation more clear. Before integration over $n_{a}$
we can use constraint $n_{a}n_{a}=r$ to put a factor $n_{a}n_{a}$
before the $\sigma$ term in Lagrangian,
\[
\mathcal{L}=\frac{1}{2}\big[\left(\partial_{\mu}n^{a}\right)^{2}+\bar{\psi}_{a}\left(i{\not\!\partial}-\sigma\right)\psi_{a}-\sigma^{2}(n_{a})^{2}-D\left(\left(n_{a}\right)^{2}-r\right)-\bar{\chi}\psi_{a}n_{a}-\bar{\psi}_{a}\chi n_{a}\big]\,.
\]
In this equation we rename the Lagrange multiplier $\la$ and call it
$D$. Thus, mass of both bosons and fermions is given by vev of the
same field $\si$ and in homogeneous vacuum state $D=0$ corresponds
to unbroken supersymmetry. The effective action 
\begin{equation}
\begin{split}
&S_{eff}=\frac{i(N\!-\!1)}{2}\,{\rm Tr}\log\left(-\partial^{2}-D-\sigma^{2}\right)-\frac{i(N\!-\!1)}{2}\,{\rm Tr}\log\left(i\!\not\!\not\!\partial-\sigma\right)\\
&\qquad\quad +\frac{1}{2}\int d^{2}x\left[\left(\partial n\right)^{2}-\left(\sigma^{2}+D\right)n^{2}+rD\right]\label{eq:susy2}\,.
\end{split}
\end{equation}
The first form of effective action (\ref{eq:susy1}) shows that fermionic
part of the model is nothing but the Gross-Neveu model (with the number
of degrees of freedom reduced by factor 2 because Majorana fermions
are used instead of Dirac ones).

From identity $\gamma^{5}\left(i{\not\!\partial}-\sigma\right)\gamma^{5}=-\left(i{\not\!\partial}+\sigma\right)$
we can obtain  
\[
{\rm Tr}\log\left(i{\not\!\partial}-\sigma\right)=\frac{1}{2}\,{\rm Tr}\log\left(-\left(i{\not\!\partial}-\sigma\right)\left(i{\not\!\partial}+\sigma\right)\right)=\frac{1}{2}\,{\rm Tr}\log\left(\partial^{2}+\sigma^{2}-i\gamma^{\mu}\partial_{\mu}\sigma\right)
\]
If $\sigma$ does not depend on time we have 
\begin{equation}
{\rm Tr}\log\left(i{\not\!\partial}-\sigma\right)=\frac{1}{2}\,{\rm Tr}\log\left(\partial^{2}+\sigma^{2}+\partial_{x}\sigma\right)+\frac{1}{2}\,{\rm Tr}\log\left(\partial^{2}+\sigma^{2}-\partial_{x}\sigma\right)\label{eq:determinants}
\end{equation}
If $\sigma$ is a topologically non-trivial solution for the GN
model, then $\lambda=\sigma^{2}\pm\partial_{x}\sigma$ is solution
to $\mathbb{CP}^{\,N-1}$ model and, thus, to $\mathbb{O}^{N}$ model.
In terms of $D$ it means $D=\pm\partial_{x}\sigma$. For definiteness
we set $\lambda=\sigma^{2}-\partial_{x}\sigma$. Thus, 
\[
S_{eff}=\frac{i(N\!-\!1)}{4}\,{\rm Tr}\log\left(-\partial^{2}\!-\!\sigma^{2}\!+\!\partial_{x}\sigma\right)-\frac{i(N\!-\!1)}{4}\,{\rm Tr}\log\left(-\partial^{2}\!-\!\sigma^{2}\!-\!\partial_{x}\sigma\right)+\frac{r}{2}\int\! d^{2}xD\,.
\]
Here we used the fact that $n$ is zero mode and that overall sign
of expression under the logarithm is unimportant because leads only
to pure imaginary constant contribution. The simplest inhomogeneous
solution 
\begin{equation}
\sigma=m\tanh mx\label{eq:kink}
\end{equation}
leads to $\la$ in form (\ref{eq:line cond}). For this solution $\sigma^{2}+\partial_{x}\sigma=m^{2}$
, so we can
see that one of two terms in (\ref{eq:determinants}) is just a vacuum
determinant and does not change energy. It is consistent with the
fact that the GN energy  ($E=Nm/2\pi$ instead of $E=Nm/\pi$ as
in Ref.\,\cite{Feinberg} because we consider Majorana fermions)
kink is minus half of energy of $\mathbb{O}^{N}$ soliton. Difference
of signs of energies can be formally explained by the different signs
of logarithms of bosonic and fermionic determinants.

\subsection{Supersymmetric \boldmath{$\mathbb{CP}^{\,N\!-\!1}$} model}

Calculation of effective action in supersymmetric $\mathbb{CP}^{\,N\!-\!1}$
model is similar to the case of supersymmetric $\mathbb{O}^{N}$
model. Supersymmetric modification of (\ref{eq:cp}) is 
\[
\mathcal{L}=D^{\mu}\bar n_{a}D_{\mu}n^{a}+\bar{\psi}_{a}i{\not\! \!D}\psi^{a}+\frac{1}{4\,r}\left(\bar{\psi}_{a}\psi^{a}\right)^{2}+\frac{1}{4\,r}\left(\bar{\psi}_{a}i\gamma^{5}\psi^{a}\right)^{2}
\]
where $\psi^{a}$ are Dirac spinors. The constraints are: $\bar n_{a} n^{a}=r$,
$\bar n_{a}\psi^{a}=0.$ We introduce Lagrange multipliers $\lambda$
and $\chi$ and auxiliary fields $\sigma$ and $\pi$,
\[
\mathcal{L}=D^{\mu}\bar n_{a}D_{\mu} n^{a}\!+\!\bar{\psi}_{a}\left(i\not\!\!{D}\!-\!\sigma\!-\!i\gamma^{5}\pi\right)\psi^{a}\!-\!r\sigma^{2}\!-\! r\pi^{2}\!-\!\lambda\left(\bar n_{a}n^{a}\!-\!r\right)\!-\!\bar{\chi}\psi^{a}\bar n_{a}\!-\!\bar{\psi}_{a}\chi n^{a}\,.
\]
The effective action is (we again set $A_{\mu}=0$) 
\[
\begin{split}
S_{eff}= \,&i(N\!-\!1){\rm Tr}\log\left(-\partial^{2}\!-\!\lambda\right)\!-\!i(N\!-\!1){\rm Tr}\log\left(i\hat{D}\!-\!\sigma\!-\!i\gamma^{5}\pi\right)\\
&+
\int\! d^{2}x\left(\partial^{\mu} \bar n \partial_{\mu}n\!-\!\lambda(|n|^{2}\!-\!r)-\sigma^{2}\!-\!\pi^{2}\right).
\end{split}
\]
Fermionic part of the action coincides with the chiral Gross-Neveu model.
This model has a continuous U(1) spontaneously broken
symmetry, so does not possess topologically stable kinks. However,
it has inhomogeneous solution (\ref{eq:kink}) and $\pi=0$ which
is stabilized by trapped fermions. For this solution the bound state
should be half-filled, see \cite{Feinberg3}. So we have
found a solution of the same type as in case of $\mathbb{O}^{N}$
model.

\section{Periodic inhomogeneous solution}

In this Section we analyze periodic solution, which corresponds to
the kink crystal in Gross-Neveu model. We explicitly check that the
gap equation is true for this solution. However, the amplitude
of the $n^{2}$ condensate has an infrared divergence. We calculate
the energy of this solution and find that it is lower than for homogeneous
solution.

\subsection{Gap equation}

In this section we check self-consistency of periodic solution. In
this calculation we follow the ideas from \cite{Thies} and use results
from \cite{Li}. For this purpose we consider possible solution $\lambda=\sigma^{2}-\partial_{x}\sigma$,
where 
\begin{equation}
\sigma=\nu m\,\frac{{\rm sn}\left(mx;\,\nu\right){\rm cn}\left(mx;\,\nu\right)}{{\rm dn}\left(mx;\,\nu\right)}
\label{eq: GN condensate}
\end{equation}
is proportional to $\bar{\psi}\psi$ condensate in the GN model. It is also possible to write this condensate in form
\begin{equation}
\sigma=m\,\frac{2\sqrt{\nu_{1}}}{1+\sqrt{\nu_{1}}}\, {\rm sn}\Big(\frac{2mx}{1+\sqrt{\nu_{1}}};\,\nu_{1}\Big),
\label{eq: GN condensate1}
\end{equation}
where parameters are connected as
\begin{equation}
\nu=\frac{4\sqrt{\nu_{1}}}{\left(1+\sqrt{\nu_{1}}\right)^{2}}.\label{eq:par}
\end{equation}
Note that solutions $\lambda=\sigma^{2}\pm\partial_{x}\sigma$ are
different only by shift on a half of period, so we do not need to
consider the solution with plus sign. For simplicity 
we will use only form \eqref{eq: GN condensate} and omit the second
argument of elliptic functions. Standard calculation yields 
\[
\lambda=m^{2}\nu\left(2\,{\rm sn}^{2}\left(mx\right)-1\right).
\]

We need to find eigenfunctions of the operator $-\partial_{x}^{2}+\lambda$.
For the operator $-\partial_{y}^{2}+2\nu\, {\rm sn}^{2}y$ (where $y=mx$)
eigenfunctions are found in \cite{Li}:
\begin{equation}
\begin{split}
&\left(-\partial_{y}^{2}+2\nu\, {\rm sn}^{2}y\right)f=\mathcal{E}f\,;\\[2mm]
&f\left(y\right)=\frac{\theta_{1}\left(\frac{\pi(y+\alpha)}{2K}\,,\,q\right)}{\theta_{4}\left(\frac{\pi y}{2K}\,,\,q\right)}\exp\left(-yZ\left(\alpha\right)\right),\quad q=\exp\left(-\pi K^{\prime}/K\right).
\end{split}
\end{equation}
Here and later $K$ and $E$ denote full elliptic integrals of the
first and the second kinds with argument $\nu$, if it is not stated
otherwise, and $K^{\prime}\left(\nu\right)=K\left(1-\nu\right).$
The parameter $\alpha=K+i\eta$ for the lower band with eigenvalues
$\nu<\mathcal{E}<1$ and $\alpha=i\eta$ for the band $\mathcal{E}>1+\nu$.
The eigenvalue can be expressed via parameter $\alpha$ as
\[
\mathcal{E}=\nu+\omega^{2}/m^{2}={\rm dn}^{2}\alpha+\nu
\]
For the states of the spectrum $Z\left(\alpha\right)$is purely imaginary
and does not change the absolute value of $f$. Using the identities for the
product of two theta-functions we can obtain
\[
\left|f\left(x\right)\right|^{2}=A^{2}\Big(1-\frac{{\rm cn}^{2}mx}{{\rm cn}^{2}\alpha}\Big).
\]
We need to fix the normalization factor $A$. The normalization condition
is that the average of the square of the eigenfunction is equal to
1,
\[
A^{2}\int_{0}^{2K/m}\Big(1-\frac{{\rm cn}^{2}mx}{{\rm cn}^{2}\alpha}\Big)dx=\frac{2K}{m}\,.
\]
The integral can be readily computed and we find normalized eigenfunctions
\begin{equation}
\left|f_{k}\right|^{2}=\frac{\omega^{2}/m^{2}-{\rm dn}^{2}mx}{\omega^{2}/m^{2}-E\left(\nu\right)/K\left(\nu\right)}\,.
\label{eq: periodic mode square}
\end{equation}
Note that for upper band both numerator and denominator are negative. 

It is convenient to integrate over the eigenvalue $\omega$ instead
of momentum $k$. To change the variable of integration, we use the
formula from \cite{Li}
\[
\frac{1}{m}\frac{dk}{d\mathcal{E}}=\frac{\nu+E/K-\mathcal{E}}{\sqrt{\left(1-\mathcal{E}\right)\left(\mathcal{E}-\nu\right)\left(1+\nu-\mathcal{E}\right)}}\,.
\]
Therefore,
\[
\frac{dk}{d\omega}=\frac{E/K-z^{2}}{\sqrt{\left(1-\nu-z^{2}\right)\left(1-z^{2}\right)}}\,,\qquad z=\omega/m\,.
\]
The gap equation can be rewritten as 
\[
n^{2}=r-\frac{N}{2\pi}\int\frac{dz}{z}\left|\frac{dk}{d\omega}\right|\left|f_{k}\right|^{2}.
\]
Integration over $z$ is over both bands. Bare coupling constant can
be expressed as 
\[
r=\frac{N}{4\pi}\int dk\bigg\{\frac{1}{\sqrt{k^{2}+\Lambda^{2}}}-\frac{1}{\sqrt{k^{2}+M^{2}}}\bigg\}=
\frac{N}{2\pi}\log \frac{M}{\Lambda}\,,
\]
where $\Lambda$ is the mass scale of the theory and $M$ is the Pauli-Villars UV cut-off. Explicit form of gap equation is 
\[
\begin{split}
&n^{2}=\frac{N}{2\pi}\log\frac{m}{\Lambda}+\frac{N}{2\pi}\int_{1}^{\infty}dz\left\{ \frac{1}{\sqrt{z^{2}-1}}-\frac{1}{z}\frac{z^{2}-
{\rm dn}^{2}mx}{\sqrt{\left(z^{2}-1+\nu\right)\left(z^{2}-1\right)}}\right\} \\
&-\frac{N}{2\pi}\int_{0}^{\sqrt{1-\nu}}\frac{dz}{z}\frac{{\rm dn}^{2}mx-z^{2}}{\sqrt{\left(1-\nu-z^{2}\right)\left(1-z^{2}\right)}}=\frac{N}{2\pi}\left(a+b\cdot {\rm dn}^{2}mx\right)
\end{split}
\]
Here we extracted the term, proportional to the square of the zero
mode of potential $\lambda$ 
\[
\psi_{0}\sim {\rm dn}\left(mx\right).
\]
The second gap equation is 
\[
\left(-\partial_{x}^{2}+\lambda\right)n=0\,,
\]
so $n$ must be proportional to  zero mode. It means that  $a=0$ and this
condition determines the parameter $m$. 

From the expressions above
we obtain 
\begin{equation}
a=\log\frac{m}{\Lambda}\!+\!\int_{1}^{\infty}\!\!\!\!dz\bigg\{ \frac{1}{\sqrt{z^{2}\!-\!1}}\!-\!\frac{z}{\sqrt{\left(z^{2}\!-\!1\!+\!\nu\right)\left(z^{2}\!-\!1\right)}}\bigg\} \!+\!\int_{0}^{\sqrt{1-\nu}}\!\!\!\!\!dz\,\frac{z}{\sqrt{\left(1\!-\!\nu\!-\!z^{2}\right)\left(1\!-\!z^{2}\right)}}\,,
\label{eq: periodic constant term}
\end{equation}
\begin{equation}
b=\int_{1}^{\infty}\frac{dz}{z}\frac{1}{\sqrt{\left(z^{2}\!-\!1\!+\!\nu\right)\left(z^{2}\!-\!1\right)}}-\int_{0}^{\sqrt{1-\nu}}\frac{dz}{z}\frac{1}{\sqrt{\left(1\!-\!\nu\!-\!z^{2}\right)\left(1\!-\!z^{2}\right)}}\,.\label{eq:periodic condensate amplitude}
\end{equation}
All the integrals are elementary functions and their calculation
is straightforward. However, the last integral in expression for $b$
is divergent in infrared. So we introduce a very small cut-off $\epsilon=\omega_{min}/m$.
Physically it corresponds to placing the system in a box of large
but finite size $L$ and dropping out zero mode from the gap equation.
Then, 
\[
k_{min}=\frac{2\pi}{L},\quad\omega_{min}=k_{min}\frac{d\omega}{dk}\left(\omega=0\right)=\frac{2\pi}{L}\sqrt{1-\nu}\,
\frac{K}{E}\,.
\]
The calculation yields 
\begin{equation}
a=\log\frac{m}{\Lambda}+\log\left(1+\sqrt{1-\nu}\right)=0\,,\qquad
m=\frac{\Lambda}{1+\sqrt{1-\nu}}\,.
\label{eq:mass periodic}
\end{equation}
Here we recall the transformation of elliptic parameter (\ref{eq:par})
and return to the original parameter $\nu$,
\[
\Lambda=\frac{2m}{1+\sqrt{\nu}}\,.
\]
Thus, the fermionic condensate can be written in the form \eqref{eq: GN condensate1},
\[
\sigma=\sqrt{\nu_{1}}\Lambda\, {\rm sn}\left(\Lambda x;\,\nu_{1}\right)\,.
\]
In terms of mass of particle in homogeneous phase this expression
takes especially simple form. However, physical reason for this simplification
is unclear.

The second coefficient 
\[
b=\frac{1}{\sqrt{1-\nu}}\log\left(\frac{1+\sqrt{1-\nu}}{Lm}\,\frac{\pi K}{E}\right).
\]
Note that this coefficient has logarithmic divergence and is negative
at sufficiently large length. It implies the inequality $n^{2}<0$.

\subsection{Energy density}

If we ignore the infrared divergence, average energy density can be
calculated in much similar way to the calculation of the energy of
soliton. Omitting rather tricky technical details we give here
the final
result is
\begin{equation}
\epsilon=\frac{N\Lambda^{2}}{4\pi}-\frac{E\left(\nu\right)}{K\left(\nu\right)}\frac{Nm^{2}}{\pi}.\label{eq:periodic energy}
\end{equation}

Now we discuss some arguments connected with calculation of energy-momentum
tensor \eqref{eq:time energy-momentum}.
Due to conservation of momentum $\partial_{\mu}\theta^{\mu}_{\nu}=0$ we have
$\partial_{x}\langle\theta_{11}\rangle=0$. The $r$ term and $n^{2}$
term cancel each other similarly to the case of soliton. The mass
term contribution
\[
\sum_{i}C_{i}m_{i}^{2}\left|n_{i}\right|^{2}=N\int\frac{dk}{2\pi}\sum_{i}\frac{C_{i}m_{i}^{2}}{2\sqrt{\omega_{k}^{2}+m_{i}^{2}}}\left|f_{k}\right|^{2}=\frac{N}{2\pi}\left(\alpha+\beta \,{\rm dn}^{2}mx\right),
\]
where the square of the mode is given by \eqref{eq: periodic mode square}.
We are going to calculate only the coefficient $\beta$,
\begin{align*}
\beta =&-\int_{1}^{\infty}dz\sum_{i}\frac{C_{i}m_{i}^{2}}{\sqrt{\left(z^{2}+a_{i}^{2}\right)\left(z^{2}-1\right)\left(z^{2}-1+\nu\right)}}\\
 & +\int_{0}^{\sqrt{1-\nu}}dz\sum_{i}\frac{C_{i}m_{i}^{2}}{\sqrt{\left(z^{2}+a_{i}^{2}\right)\left(z^{2}-1\right)\left(z^{2}-1+\nu\right)}}=-m^{2}.
\end{align*}
We are not able to calculate
derivative terms in energy-momentum tensor but the fact that $\langle\theta_{11}\rangle=const$
suggests that 
\[
\sum_{i}C_{i}\left(\left|\partial_{t}n_{i}\right|^{2}+\left|\partial_{x}n_{i}\right|^{2}\right)=\frac{N}{2\pi}\left(\alpha_{1}+\beta\,{\rm dn}^{2}mx\right)
\]
with the same coefficient $\beta$ but different coefficient $\alpha_{1}$.
Therefore energy density is
\[
\epsilon\left(x\right)=\langle\theta_{00}\rangle=-\frac{Nm^{2}}{\pi}{\rm dn}^{2}mx+const.
\]
This result is consistent with the formula 
\[
\epsilon\left(x\right)=\frac{N}{2\pi}\,\lambda\left(x\right)+const.
\]
The value of the constant can be determined from the average energy
density 
\[
\epsilon\left(x\right)=\frac{N}{2\pi}\lambda\left(x\right)-\frac{N\Lambda^{2}}{4\pi}\left(\frac{1-\sqrt{1-\nu}}{1+\sqrt{1-\nu}}\right).
\]

The obtained energy is lower than the one of homogeneous solution.
However, due to infrared divergence this solution can possibly be
considered on a finite part of a plane only.

\subsection{\boldmath{$n\,\lambda$} cross-term correction}

In the subsection we compute effective action more carefully, taking
into account the quadratic quantum $n_{q}\lambda_{q}$ terms in the Lagrangian. 
Such term is absent in the standard analysis in the confinement phase.
For simplicity we consider
$\mathbb{O}^{\,N}$ model  and suppose that similar results are
valid for the $\mathbb{CP}^{\,N\!-\!1}$ model. We find out that additional term
is a $1/N$ correction to the effective action and therefore should
not be taking into consideration to the leading order. 

The partition function is
\begin{equation}
Z=\int DnD\lambda\exp\left\{ -S\right\} ,\label{eq:partfunc}
\end{equation}
where the action is obtained from \eqref{eq:on} after proper rescaling
\begin{equation}
S=\frac{1}{2}\int d^{2}x\left(\left[\partial n\right)^{2}+\lambda\left(n^{2}-r\right)\right].\label{eq:action}
\end{equation}
We separate the fields into classical and quantum components,
\begin{equation}
n=n_{cl}+n_{q},\quad n_{cl}=\left(n_{0},0,\dots,0\right),
\quad
\lambda=\lambda_{0}+\lambda_{q}\,,\label{eq:l field}
\end{equation}
and perform
functional integration over the quantum components in the Gaussian
approximation. 
The action in terms of quantum and classical components
\begin{equation}
S=\frac{1}{2}\int d^{2}x\left[\left(\partial n_{0}\right)^{2}+\lambda_{0}\left(n_{0}^{2}-r\right)+\left(\partial n_{q}\right)^{2}+\left(\lambda_{0}+\lambda_{q}\right)n_{q}^{2}+2n_{0}\lambda_{q}n_{1q}\right].\label{eq:action full}
\end{equation}
After integration over all but the first components of the $n_{q}$ fields
we obtain effective action,
\begin{equation}
S_{eff}^{(1)}=\frac{N\!-\!1}{2}\,{\rm Tr}\log\left(-\partial^{2}\!+\!\lambda_{0}\!+\!\lambda_{q}\right)+\frac{1}{2}\!\int \!\!d^{2}x\left[\left(\partial n_{0}\right)^{2}\!+\!\lambda_{0}\left(n_{0}^{2}\!-\!r\right)\!+\!n_{1}\!\left(-\partial^{2}\!+\!\lambda_{0}\right)n_{1}\!+\!2n_{0}\lambda_{q}n_{1}\right]\!.\label{eq:eff1}
\end{equation}
To deal with the cross term $\lambda\, n$ we shift the variable of
functional integration and obtain Gaussian integrals for $n$ and $\lambda$
\begin{equation}
\begin{split}
&n_{1}\to n_{1}+\chi\,,\;~~\chi=-\frac{1}{-\partial^{2}+\lambda}\,n_{0}\lambda_{q}\,,\\
&n_{1}\left(-\partial^{2}+\lambda_{0}\right)n_{1}+2n_{0}\lambda_{q}n_{1}\to n_{1}\left(-\partial^{2}+\lambda_{0}\right)n_{1}-n_{0}\lambda_{q}\,\frac{1}{-\partial^{2}+\lambda_{0}}\,n_{0}\lambda_{q}\,.
\end{split}
\label{eq:shift}
\end{equation}
Integration over $n_{1q}$ is trivial. However, effective action for $\lambda$
contains a complicated integral operator K with the kernel $K\left(x,\,y\right)$.
This kernel can be expressed in terms of the Green function of the
$n$ field in the $\lambda_{0}$ background,
$G\left(x,\,y\right)=\langle x|(-\partial^{2}+\lambda_{0})^{-1}|y\rangle$,  
\begin{equation}
\begin{split}
S_{eff}^{(2)}=\frac{N\!-\!1}{2}\,{\rm Tr}\log\left(-\partial^{2}+\lambda_{0}\right)-\frac{N\!-\!1}{4}\,{\rm Tr}\left(\frac{1}{-\partial^{2}+\lambda_{0}}\lambda_{q}\right)^{2}+\\
\frac{1}{2}\int d^{2}x\left[\left[\partial n_{0}\right)^{2}+ \lambda_{0}\left(n_{0}^{2}-r\right)-n_{0}\lambda_{q}\frac{1}{-\partial^{2}+\lambda_{0}}n_{0}\lambda_{q} \right]. 
\end{split}
\label{eq:eff2}
\end{equation}
The action for the $\lambda_{q}$ reads as
\begin{equation}
S_{\lambda}\!=\!-\frac{N\!-\!1}{4}\!\int \!\!d^{2}xd^{2}yG\left(x,\,y\right)G\left(y,\,x\right)\lambda_{q}\left(x\right)\lambda_{q}\left(y\right)-\frac{1}{2}\!\int\!\! d^{2}xd^{2}y\lambda_{q}\left(x\right)\lambda_{q}\left(y\right)n_{0}\left(x\right)n_{0}\left(y\right)G\left(x,\,y\right),\label{eq: l action}
\end{equation}
\[
S_{\lambda}=-\frac{1}{2}\int d^{2}xd^{2}y\lambda_{q}\left(x\right)\lambda_{q}\left(y\right)K\left(x,\,y\right),
\]
where the kernel is
\begin{equation}
K\left(x,\,y\right)=\frac{N\!-\!1}{2}\,G\left(x,\,y\right)^{2}+n_{0}\left(x\right)n_{0}\left(y\right)G\left(x,\,y\right).\label{eq:kernel}
\end{equation}
The final answer for the effective action is 
\begin{equation}
S_{eff}=\frac{N\!-\!1}{2}\,{\rm Tr}\log\left(-\partial^{2}+\lambda_{0}\right)+\frac{1}{2}\,{\rm Tr}\log K+\frac{1}{2}\int d^{2}x\left[\left(\partial n_{0}\right)^{2}+\lambda_{0}\left(n_{0}^{2}-r\right)\right].\label{eq:eff action}
\end{equation}
The second term in the effective action is the correction we have
calculated. In this expression all terms but the second contain a
large $N$ factor. So the correction is suppressed in large $N$ limit.

\subsection{Comment on GN model at zero density}
For comparison let us briefly comment on the periodic solution in Gross-Neveu
model with the Minkowski Lagrangian
\[
\mathcal{L}=\bar{\psi}\left(i\!\!\not\!{\partial}-\sigma\right)\psi-r\,\sigma^{2}.
\]
The similar problem was considered in \cite{Thies}. For more similarity,
in this section we consider the theory with Dirac fermions.
Generically the period of the elliptic solution to the
GN model is fixed by the chemical potential however for the
zero density case  we do not have the Fermi momentum parameter, the period
of the solution remains a free parameter.

The effective action is 
\[
S_{eff}=-iN{\rm Tr}\log\left(i\!\!\not\!{\partial}-\sigma\right)-r\int d^{2}x\sigma^{2}.
\]
We look for the solution in the form \eqref{eq: GN condensate}. The
mass parameter $m$ of this solution is connected to the mass scale
$\Lambda$ of the theory through the gap equation that reads as
\[
\sigma\left(x\right)=\frac{N}{2r}\int\frac{dk}{2\pi}\,\bar{\psi}_{k}\psi_{k}\,,
\]
where eigenfunctions
\[
\bar{\psi}_{k}\psi_{k}=\frac{\omega}{\omega^{2}-m^{2}E/K}\,\sigma\left(x\right).
\]
Therefore gap equation reduces to
\[
1=\frac{N}{2r}\int\frac{dk}{2\pi}\,\frac{\omega\left(k\right)}{\omega^{2}\left(k\right)-m^{2}E/K}\,.
\]

The fermionic gap equation leads to the same formula \eqref{eq:mass periodic}
for mass as bosonic one. Note that there is no infrared divergence.
The energy of this solution can be calculated from the relation \eqref{eq:determinants}
between bosonic and fermionic determinants. Using the fact that the
potentials $\sigma^{2}\pm\partial_{x}\sigma$ we find that energy
density for fermionic case is different from bosonic only by sign,
\[
\epsilon_{\,GN}=-\epsilon=-\frac{N\Lambda^{2}}{4\pi}+\frac{E\left(\nu\right)}{K\left(\nu\right)}\frac{Nm^{2}}{\pi}\,.
\]
Thus, the energy is minimal for homogeneous solution which is the correct ground state.
However, the non-vanishing chemical potential modifies the ground state which becomes 
inhomogeneous.

\section{Discussion}

In this paper we considered the properties of the inhomogeneous solutions  \cite{Nitta}
found recently for $\mathbb{CP}^{\,N\!-\!1}$ sigma-model at large $N$.
We focused at the  soliton-like solution and the elliptic solution
to the quantum gap equation. 
The careful analysis shows that
the energy of the soliton is lower than the energy of the homogeneous ground
state. This clearly makes questionable the common viewpoint that the ground
state of the $\mathbb{CP}^{\,N\!-\!1}$ sigma-model at large $N$ is homogeneous.

The answer to the question about  the true ground state of the model does not look simple.
The na\"{i}ve conjecture would be that the 
periodic elliptic kink crystal solution yields the true ground state and vacuum is in FFLO-like phase
as in GN model with non-vanishing chemical potential.
The energy for kink crystal solution can be evaluated and indeed it is lower than
energy of the homogeneous state. However there are two points of concern which 
provide the difficulties with  such immediate  identification. First,the kink crystal solution 
suffers from the IR divergence at the infinite plane and deserves some 
IR regularization, for instance by introducing a box. Secondly the kink crystal solution has the 
free massive parameter which fixes the period whose interpretation is not completely clear
in non-SUSY case.
It is counterpart of the chemical potential in the GN model.

It is instructive to look at the massive deformations of the large $N$ sigma-models.
It has been discussed in \cite{ferrari} for  $\mathbb{O}^{N}$  and in  \cite{gsy} for $\mathbb{CP}^{\,N\!-\!1}$.
The mass 
provides the IR regularization of the models, at large masses the theory 
can be treated perturbatively and is proven to be in the Higgs-like phase. 
In both models there is a clear-cut  phase transition 
at the value of the mass of order of nonperturbatively generated scale $\Lambda$.
Moreover, it is demonstrated in \cite{ferrari} that at the phase transition point two states
become massless: the bound state of two n-particles and the soliton. 

For masses below $\Lambda$ these light states could hint at existence of 
a dual, more  suitable, description. This is similar to the Sine-Gordon model transition 
from the bosonic description at weak coupling to the fermionic one at strong coupling.
We did not explore this opportunity. Instead, in our analysis we suggest 
 that the  ground
state of these models is a small mass deformation of the FFLO-like 
kink crystal solution. The (twisted) mass parameter fixes the period of the
elliptic solution to the gap equation and provides the IR regularization
hence everything is well defined in this case.
We hope to investigate this issue elsewhere. 

The massive deformations of the 2D theories have the clear-cut 4D counterparts --
these are the gauge theories with flavor and  masses of fundamental matter
play the similar role. Instead of the kinks in 2D the domain walls in 4D
are considered and the nontrivial mass dependence of their tensions are of
interest. We would like to mention two examples: QCD at $\theta=\pi$ and 
softly broken ${\cal N}=2$ SQCD. In both cases there are domain walls with
mass dependent tensions. In QCD case it was proved in \cite{gks} that 
the 3D theory on the domain wall is deconfined. However, the approach of \cite{gks} 
does not give exactly the critical value of the quark mass when 
the domain wall tension vanishes.
On the other hand, in softly broken
${\cal N}=2$ SQCD at $N_f=1$ the critical value of the mass at the
Argyres-Douglas point when the domain wall tension vanishes has been found 
exactly \cite{gvy}. At the critical mass the whole 4D  theory turns out in the
deconfinement phase \cite{gvy} and this fits with the deconfinement 
in 3D theory on the domain wall observed in \cite{gks}. Indeed, when the domain
wall tension is small it becomes wide and finally the deconfined phase
occupies the whole space-time at the Argyres-Douglas point.

One more comment is in order. Recently, it was recognized that the discrete 
anomaly matching provides the powerful tool for the analysis of the phase
diagram of the strongly coupled theories. In particular this approach has
been applied to the discussion of the ground state  
in the spin systems with the ${\rm SU}(N)$ structure group in some representation
\cite{tim}. As was known for a while \cite{haldane} that the low-energy action
for the ${\rm SU}(2)$ group case gets identified with the $\mathbb{CP}^{\,\!1}$ model
with the $\theta$ term which depends on the spin representation. If $\theta=\pi(2k+1)$ 
the ground state turns out to be gapless and can be thought of as the 
the condensate of dimers. More recent analysis \cite{tim} suggests that the similar
gapless phases for higher spin chains could occur at $\theta=2\pi/N$.
For instance, in ${\rm SU}(3)$ case at proper value of $\theta$  the ground state is gapless 
and presumably is a kind of condensate
of trimers. We could speculate that  gapless ground state we have found
could be some analogue of the Haldane's gapless phase and our periodic kink crystal
is the generalization of the dimer and trimer condensates ground states for low rank
spin systems. Indeed our soliton-like solution from the chiral GN viewpoint
can be considered as the superposition of $N$ elementary kinks in the 
hedgehog shape. In our case we have $\theta=0$ but presumably it can be reasonable 
approximation of  $\theta=2\pi/N$ at large $N$.   

We have touched a bit the SUSY generalization of the new solution  postponing
the detailed analysis for the separate study. The immediate question
concerns the BPS property of the solution. The SUSY picture implies also
the several questions concerning its brane interpretation. Let us make a few remarks

\begin{itemize}
\item The nontrivial profile of the n-field corresponds to the pulling of
D2 brane in particular direction by D2-D4 string. Hence to some extent the soliton
is represented by the profile of F1 D2-D4 string. It would be interesting to 
get the interpretation of the soliton solution from the F1 worldsheet viewpoint  

\item The brane picture for the GN model \cite{kutasov} tells that the
kink corresponds to the interpolation between two possible intersections 
of D4 and D6 branes.
This resembles the appearance of the second vacuum in the $\mathbb{CP}^{\,N\!-\!1}$ model
coupled to 4D degrees of freedom \cite{gukov}. Hence it is natural
to expect that the brane configuration responsible for the soliton and
soliton lattice configurations involves D6 branes.

\item The local negative energy contribution is typical for boojoums \cite{rev1}  when
the magnetic non-abelian string is attached to the domain wall. The negative energy is
localized on the domain wall near the intersection point. One could conjecture
that the soliton solution corresponds to the region of the intersection 
of the D6 domain wall and D2 brane representing non-abelian string in 4D gauge theory. 

\item Recently the so-called negative branes with the negative tensions have been
found \cite{vafa}. These objects are identified both  for extended branes and for 
point-like particles with negative mass. For some of them the supergravity solutions have been found
and it was argued that they obey the fermion statistics. It is unclear if our 
finding is related with this issue.

\end{itemize}

Several questions concerns the IR properties of the periodic solution.

\begin{itemize}
\item Connection between infrared divergences in the 
solution and Coleman's theorem deserves the careful study. There
are some example of models in which 2D continuous symmetry can be
broken (chiral GN and $\mathbb{CP}^{\,N\!-\!1}$ on a circle at large $N$, SUSY $\mathbb{CP}^{\,N\!-\!1}$
due to mixing of $\pi$ and $A_{\mu}$ propagators). Could something
similar happen in our case?

\item Our study imply that the homogeneous solution for $\mathbb{CP}^{\,N\!-\!1}$ model
certainly is not the true ground state contrary to the standard viewpoint. 
Therefore it is necessary to clarify if it the the metastable
minimum of just local extremum. If it is the metastable state
the kink-antikink configuration usually considered as the
excitation could be treated as the bounce responsible for the
decay of the metastable vacuum.

\item Even if periodic solutions do not exist on a plane, they can change
phase structure on a circle. There are possible phase transitions
when $n^{2}=0$.
\end{itemize}

Let us remark that the lattice studies of the $\mathbb{CP}^{\,N\!-\!1}$ model also shows a  unexpected 
structure of the ground state \cite{thacker} which has in the Euclidean space  
the crystal-like double-layer structure.
The distribution of the topological charge density has the dipole-like
structure and vacuum was interpreted as a kind of condensate of the 
Wilson loops. It is unclear if the kink crystal solution we considered in this study 
with minimal energy has something to do with these lattice observations.

\section*{Acknowledgements}
We are grateful to I. Klebanov, N. Nekrasov and M. Shifman for the useful discussions. 
We appreciate hospitality of the Kavli Institute for Theoretical Physics at UCSB during the programs
``Resurgent Asymptotics in Physics and Mathematics'' and ``Knots and Supersymmetric Gauge Theories'' 
where our research was supported  in part by the National Science Foundation under Grant No.\ NSF PHY-1748958. 
The work of A.G. was partially supported by the Basis Foundation Fellowship. A.G. also thanks the Simons 
Center for Geometry and Physics for the  hospitality and support during the program ``Exactly Solvable Models of Quantum Field Theory and Statistical Mechanics.''

\begin{appendices}

\section{Effective action calculation for soliton\label{sec:Effective-action-calculation}}

Here we provide the technical details of computation of energy of
the soliton. The coupling constant can be found from the gap equation
for the homogeneous solution in space of large volume $V$,
\begin{equation}
\begin{split}
&r\cdot V=\sum_{i=0}^{I}C_{i}\,{\rm Tr}\,\frac{1}{-\partial^{2}+m_{i}^{2}+m^{2}}=V\cdot\int\frac{d^{2}k}{4\pi^{2}}\sum_{i=0}^{I}C_{i}\,\frac{1}{k^{2}+m_{i}^{2}+m^{2}},\\
&r=-\frac{N}{4\pi}\sum_{i=0}^{I}C_{i}\log\left(m^{2}+m_{i}^{2}\right).\label{eq:coupling PV}
\end{split}
\end{equation}
 The trace of the operator can be written as a sum over the eigenvalues,
\[
{\rm Tr}\log\left(-\partial^{2}+m_{i}^{2}+\lambda\right)=T\int\frac{d\omega}{2\pi}\,\sum_{n}\log\left(\omega^{2}+\omega_{n}^{2}+m_{i}^{2}\right).
\]
Here $T$ stands for a large time cut-off and summation is over all
eigenvalues $\omega_{n}^{2}$ of the operator $-\partial_{x}^{2}+\lambda$.
Therefore, we obtain the following expression for energy:
\begin{equation}
E_{1}=N\int_{-\infty}^{+\infty}\frac{d\omega}{2\pi}\sum_{n}\sum_{i=0}^{I}C_{i}\log\left(\omega^{2}+\omega_{n}^{2}+m_{i}^{2}\right)-r\!\int_{-\infty}^{+\infty}\!dx\lambda\,.
\label{eq: energy from determinant}
\end{equation}
The same expression can be written for energy of vacuum $E_{vac}$
when $\lambda=m^{2}$ and eigenvalues are $\omega_{0n}^{2}$.

We use expression \eqref{eq: energy from determinant}  for the energy 
and subtract vacuum contribution:
\begin{align*}
E & =E_{1}-E_{vac}=N\int_{-\infty}^{+\infty}\frac{d\omega}{2\pi}\sum_{i=0}^{I}C_{i}\log\left(\omega^{2}+m_{i}^{2}\right)+\\
 & +N\int_{-\infty}^{+\infty}\frac{d\omega}{2\pi}\sum_{n}\sum_{i=0}^{I}C_{i}\log\frac{\omega^{2}+\omega_{n}^{2}+m_{i}^{2}}{\omega^{2}+\omega_{0n}^{2}+m_{i}^{2}}-\int_{-\infty}^{+\infty}dx\left(\lambda-m^{2}\right)r.
\end{align*}
Here the first term is contribution from the zero mode and the second
is contribution from the continuum. If we use integral 
\[
\int_{-\infty}^{+\infty}\frac{d\omega}{2\pi}\log\left(1+\frac{a^{2}}{\omega^{2}}\right)=a
\]
and integrate over $\omega$ in the first and second term, and over
coordinate in the third we arrive at 
\[
E=N\sum_{i=0}^{I}C_{i}m_{i}+N\sum_{n}\sum_{i=0}^{I}C_{i}\left(\sqrt{\omega_{n}^{2}+m_{i}^{2}}-\sqrt{\omega_{0n}^{2}+m_{i}^{2}}\right)+4mr.
\]
Summation over all eigenvalues can be replaced with integration over
all momenta,
\[
\sum_{n}\to\int dk\rho\left(k\right),\quad \omega_{n}^{2}\to k^{2}+m^{2},
\]
where difference of densities of states for homogeneous and inhomogeneous
states is
\[
\rho\left(k\right)=\frac{1}{\pi}\frac{d\delta\left(k\right)}{dk}=-\frac{2m}{\pi\left(k^{2}+m^{2}\right)}.
\]
Here $\delta\left(k\right)=\pi-2\arctan({k}/{m})$ is phase shift
for eigenfunctions \eqref{eq:soliton eigenfunctions}. Therefore energy
is
\[
E=N\sum_{i=0}^{I}C_{i}m_{i}-\frac{2Nm}{\pi}\int_{0}^{+\infty}dk\sum_{i=0}^{I}C_{i}\frac{\sqrt{k^{2}+m^{2}+m_{i}^{2}}}{k^{2}+m^{2}}+4mr.
\]
We use integral 
\[
\int dk\frac{\sqrt{k^{2}+m^{2}+M^{2}}}{k^{2}+m^{2}}=\frac{M}{m}\arctan\frac{Mk}{m\sqrt{k^{2}+m^{2}+M^{2}}}+\log\left(k+\sqrt{k^{2}+m^{2}+M^{2}}\right)
\]
and obtain
\[
E=N\sum_{i=0}^{I}C_{i}m_{i}-\frac{2Nm}{\pi}\left[\sum_{i=1}^{I}C_{i}\frac{m_{i}}{m}\arctan\frac{m_{i}}{m}-\frac{1}{2}\sum_{i=0}^{I}C_{i}\log\left(m^{2}+m_{i}^{2}\right)\right]+4mr.
\]
If we apply the expression \eqref{eq:coupling PV} for $r$ and assume
that $m_{i}\gg m$ and thus $\arctan\left(m_{i}/m\right)=\pi/2-m/m_{i}$
we obtain 
\begin{align*}
E & =N\sum_{i=0}^{I}C_{i}m_{i}-\frac{2Nm}{\pi}\sum_{i=1}^{I}C_{i}\frac{m_{i}}{m}\frac{\pi}{2}+\frac{2Nm}{\pi}\sum_{i=1}^{I}C_{i}+\\
 & +\frac{Nm}{\pi}\sum_{i=0}^{I}C_{i}\log\left(m^{2}+m_{i}^{2}\right)-\frac{Nm}{\pi}\sum_{i=0}^{I}C_{i}\log\left(m^{2}+m_{i}^{2}\right).
\end{align*}
We see that all terms except the third cancel. The sum in the third
term is $\sum_{i=1}^{I}C_{i}=-C_{0}=-1$ and we find the expression
(\ref{eq:energy}).

\section{Energy-momentum tensor of the soliton \label{sec:Energy-momentum-tensor-of}}

To calculate the average of energy-momentum tensor components \eqref{eq:time energy-momentum}
and we need following combinations
\begin{equation}
\sum_{i}C_{i}m_{i}^{2}\langle\left|n_{i}\left(x\right)\right|^{2}\rangle=N\int\frac{dk}{2\pi}\sum_{i}\frac{m_{i}^{2}\left|f_{k}\left(x\right)\right|^{2}}{2\sqrt{k^{2}+m^{2}+m_{i}^{2}}}+N\psi_{0}\left(x\right)^{2}\sum_{i}\frac{C_{i}m_{i}}{2},\label{eq:mass}
\end{equation}
\begin{equation}
\sum_{i}C_{i}\langle\left|\partial_{x}n_{i}\left(x\right)\right|^{2}\rangle=\left(\partial_{x}n_{cl}\left(x\right)\right)^{2}+N\int\frac{dk}{2\pi}\sum_{i}\frac{\left|\partial_{x}f_{k}\left(x\right)\right|^{2}}{2\sqrt{k^{2}+m^{2}+m_{i}^{2}}}+N\psi_{0}\left(x\right)^{2}\sum_{i}\frac{C_{i}}{2m_{i}},\label{eq:spacial}
\end{equation}
\begin{equation}
\sum_{i}C_{i}\langle\left|\partial_{t}n_{i}\left(x\right)\right|^{2}\rangle=N\int\frac{dk}{4\pi}\sum_{i}C_{i}\sqrt{k^{2}+m^{2}+m_{i}^{2}}\left|f_{k}\left(x\right)\right|^{2}+N\psi_{0}\left(x\right)^{2}\sum_{i}\frac{C_{i}m_{i}}{2}.\label{eq:time}
\end{equation}
The expressions for modes and their derivatives are
\[
\left|f_{k}\left(x\right)\right|^{2}=\frac{k^{2}+m^{2}\tanh^{2}mx}{k^{2}+m^{2}}=1-\frac{m^{2}}{k^{2}+m^{2}}\frac{1}{\cosh^{2}mx},
\]
\begin{equation}
\left|\partial_{x}f_{k}\left(x\right)\right|^{2}=k^{2}+\frac{m^{2}}{\cosh^{2}mx}+\frac{m^{4}}{k^{2}+m^{2}}\left(\frac{1}{\cosh^{4}mx}-\frac{1}{\cosh^{2}mx}\right).\label{eq: mode derivative}
\end{equation}

We consider mass term \eqref{eq:mass} and terms with derivatives
\eqref{eq:spacial} and \eqref{eq:time} separately
\begin{align*}
\sum_{i}C_{i}m_{i}^{2}\langle\left|n_{i}\left(x\right)\right|^{2}\rangle & =N\int\frac{dk}{2\pi}\sum_{i}\frac{C_{i}m_{i}^{2}}{2\sqrt{k^{2}+m^{2}+m_{i}^{2}}}+\\
 & +N\frac{m^{2}}{\cosh^{2}mx}\left(-\int\frac{dk}{4\pi}\sum_{i}\frac{C_{i}m_{i}^{2}}{\left(k^{2}+m^{2}\right)\sqrt{k^{2}+m^{2}+m_{i}^{2}}}+\sum_{i}\frac{C_{i}m_{i}}{4m}\right).
\end{align*}
The first term yields the energy density of homogeneous state. Note
that in the expression \eqref{eq:spacial} for the spacial derivative
the term with derivative of classical component cancels with the convergent
part of the integral, which is a contribution from the third term
in \eqref{eq: mode derivative}. So we can write down the remaining
contributions 
\[
\sum_{i}C_{i}\langle\left|\partial_{x}n_{i}\left(x\right)\right|^{2}+\left|\partial_{t}n_{i}\left(x\right)\right|^{2}\rangle=
\]
\[
=N\frac{m^{2}}{\cosh^{2}mx}\left(\int\frac{dk}{4\pi}\sum C_{i}\left(\frac{1}{\sqrt{k^{2}+m^{2}+m_{i}^{2}}}-\frac{\sqrt{k^{2}+m^{2}+m_{i}^{2}}}{k^{2}+m^{2}}\right)+\sum_{i}\frac{C_{i}m_{i}}{4m}\right).
\]

All integrals can be computed elementary. Thus we find that contribution
to the inhomogeneous part of energy density from derivative terms
(\ref{eq:time}) and (\ref{eq:spacial}) and term (\ref{eq:mass})
with are equal. Therefore, corresponding contributions in the momentum
flaw $\theta_{11}$ in \eqref{eq:time energy-momentum} cancel and this component
does not depend on the coordinate. Combining the results, we obtain
\eqref{eq: average time} and \eqref{eq: average space}.

\end{appendices}

\end{document}